# The method to increase the thrust of high Mach number Scramjets

Yunfeng Liu[1,2] *, Xin Han[1,2], Wenshuo Zhang[1,2], Kaifu Ma[1,2]

[1] Institute of Mechanics, Chinese Academy of Sciences, Beijing 100190, China

[2] School of Engineering Science, University of Chinese Academy of Sciences, Beijing 100049, China

( Corresponding author: liuyunfeng@imech.ac.cn)

**Abstract:** The problem of engine unstart of scramjets has not been resolved. In this paper, the mechanism of engine unstart is discussed from the point of view of shock/shock interaction and deflagration-to-detonation transition. The shock/shock interaction leads to the nonlinear, transient and discontinuous process of the supersonic combustion flow field. This process is similar to the deflagration-to-detonation transition process. If the velocity of pre-combustion shock wave is faster than the velocity in the isolator, it will propagate upstream and cause the engine unstart. The C-J detonation velocity is defined as the stable operation boundary of scramjets, which is the maximum shock wave produced by combustion theoretically. The scramjets will work stable if the velocity in the isolator is faster than the corresponding C-J detonation velocity. The combustion characteristics and propulsive performance of scramjets is theoretically analyzed by using C-J detonation theory. For high Mach number scramjets, the velocity in the isolator is much faster than the C-J detonation velocity. Therefore, extra fuel and oxygen can be injected into the combustor to increase the thrust as long as the shock wave velocity driven by the combustion products is slower than the air velocity in the isolator. The theoretical results agree well with the existing experimental results, which can be used as a baseline for the development of scramjets.

**Keywords:** Scramjets, Supersonic combustion, Deflagration-to-detonation transition, unstart mechanism, thrust

# 1 Introduction

Supersonic combustion ramjet engines (Scramjets) have been expected to be one of the most promising propulsion systems for hypersonic air-breathing vehicles because they have the potential to reduce the costs of access to space by taking air from atmosphere as an oxidizer [1-4]. It also plays an important role in the reusable rocket–scramjet–rocket launch system in the near future [5]. Recent flight experiments in hypersonic vehicle technology have had some successful outcomes (ex. NASA X-43A, Boeing X-51A) , but also have seen some unexpected failures. One important failure mode is the problem of engine unstart. In engine unstart, the engine expels a flame from the inlet and loses thrust drastically.

Tsien and Beilock firstly solved the one-dimensional conservation equations by adding the source term in the energy equation to study the supersonic combustion [6]. They obtained theoretical solutions for simple one-dimensional steady diabatic flows. But the paramount question is whether the shock-free combustion can be established in steady supersonic flow. In reality, the intensive local heat release may lead to a pre-combustion shock wave in the isolator. In this case, the one-dimensional conservation equations cannot be applied to the flow with pre-combustion shock waves. If the velocity of the pre-combustion shock wave is faster than the velocity in the isolator, it will propagate out of the inlet and cause the engine unstart.

There have been many valuable papers discussing the transient process of the engine unstart during the past 60 years and we cannot include all of them here. We just introduce some typical experimental results conducted in recent years. The HyShotII scramjets experiments were conducted in the HEG shock tunnel of DLR [7-10]. The simulated flight Mach number was Mach 7.5 and the fuel was hydrogen. The velocity and static temperature at the entrance of combustor were 1720m/s and 1350K, respectively. The pre-combustion shock wave was found to begin to appear in the isolator at the equivalate ratio of ER=0.50. The engine unstart occurred at ER=1.0 and the velocity of the pre-combustion shock wave relative to the velocity in the isolator was about 300m/s in this case.

The combustion and flame stabilization modes in a hydrogen fueled scramjet combustor were investigated experimentally at the direct-connect supersonic combustion facility of China Aerodynamics Research and Development Center (CARDC) [11]. The stagnation conditions were 950 K and 0.82 MPa and the Mach number at the isolator entrance was Mach 2.0, respectively. The combustion-induced back pressure at ER=0.23 and 0.3 disturbed the flow in the isolator and the significant pressure rise points were measured in the upstream of the hydrogen injection site. The ethylene fueled scramjet combustion experiments were also conducted in this facility and similar phenomena were observed [12,13].

The rocket-based combined-cycle (RBCC) engine was tested in the ramjet engine test facility of JAXA, Japan [14]. The phenomena of engine unstart and pulsative combustion were obtained. A one-dimensional analysis was applied to the diffuser flow to identify the causes of the flow breakdown. The thermal choking condition was calculated by incorporating results of a chemical equilibrium numerical code. The fuel rates causing the flow choking were reproduced and matched with the limit of fuel rates observed in the tests. Their analysis showed that the engine would easily cause choking in the diffuser because of the larger propellant supply rates. Reasonable operation of the wind tunnel to control the flow choking were examined.

The ignition transients in a scramjet engine with air-throttling was tested at Wright-Patterson Air Force Base and corresponding numerical simulations were conducted [15-18]. The method of air-throttling at the downstream of combustor was used to promote the ignition. The airstream at the isolator entrance had a static temperature of 560 K, a static pressure of 0.328 atm, and an axial velocity of 1045 m/s, respectively. The corresponding mass flow rate was 0.757 kg/s and the Mach number 2.2. Gaseous ethylene fuel was injected into the chamber at a 15° inclination to the wall. With the increase of equivalence ratio, the pre-combustion shock wave in the isolator was observed. The instability modes were discussed including injector-flame feedback, shock-flame acoustic feedback, and shock-flame convective-acoustic feedback.

Sun et al. experimentally investigated the flame dynamics inside an ethylene-fueled scramjet combustor in a Mach 2.1 facility with the stagnation temperature of 846 K [19]. They found that when the fuel injection upstream of the cavity flameholder produces a premixed region with sufficiently high global equivalence ratio, a rapid flame flashback occurred against the incoming supersonic flow. Analysis of the experimental data suggested that the flame flashback was related to the flame acceleration, which is similar to deflagration-to-detonation transition (DDT) process. With high equivalence ratio, the immediate flame re-ignition and intense flame flashback decreased the flame blowoff duration and finally led to a more efficient heat release compared to the cases with lower equivalence ratio.

The HIFiRE scramjets with gaseous hydrogen as fuel were experimentally tested at higher flight Mach numbers in recent years and the engine unstart phenomena were improved significantly at higher equivalence ratios [20-23]. For the HIFiRE scramjet at Mach 7.5, the robust combustion was obtained in the flowpath at a range of equivalence ratios between ER=0.48 and 0.84 with a significant combustion-generated pressure-rise. The engine unstart occurred at equivalence levels close to one. Stoichiometric levels with high combustion efficiency were shown to be possible if the fuelling was shared between the inlet and the combustor. For the HIFiRE M12REST engine at flight Mach number of Mach 12, the equivalence ratio was increased to ER=1.26 without the occurrence of engine unstart [23].

Urzay gave a review on supersonic combustion in air-breathing propulsion systems for hypersonic flight and pointed out that engine unstart is a poorly understood phenomenon closely related to the dynamics of fuel injection and combustion [24]. Chang et al. reviewed the recent research progress on unstart mechanism, detection and control of hypersonic inlet [25]. A review of the unstart phenomena induced by flow choking in scramjet inlet-isolators was given by Im and Do [26]. Therefore, the study of the engine unstart mechanism is very important for the stable operation of scramjets. In this paper, we analyze the unstart mechanism theoretically by using the theory of shock/shock interaction and deflagration-to-detonation transition and propose a method to increase the thrust for high flight Mach number scramjets.

**2 Theoretical Analysis of Transient Unstart Process**

In this paper, we give a new mechanism of the transient unstart process of scramjets, which is caused by shock/shock interaction and DDT process. There are two typical structures of shock/shock interaction shown in Fig.1. The first one is the superposition of shock waves (SWs) moving in the same direction (Fig.1a). The intensive heat release in the combustor produces a series of shock waves. The latter can catch up with the former and form a new stronger shock wave. If the velocity of the new transmitted shock wave is faster than the velocity in the isolator, it will propagate out of the inlet and make the engine unstart. If this shock wave propagates in the combustible mixture, it will initiate a detonation wave. The second one is the Mach Stem (MS) caused by shock/shock interaction. The back pressure produced by supersonic combustion leads to the separation of the boundary layer. New oblique shock waves (OSWs) are formed in front of separation bubbles. The shock/shock interaction can produce a Mach Stem. Thermal choking occurs behind the Mach Stem and an over-driven detonation will be initiated. If the velocity of this over-driven detonation or this Mach Stem is faster than the velocity in the isolator, engine unstart occurs.

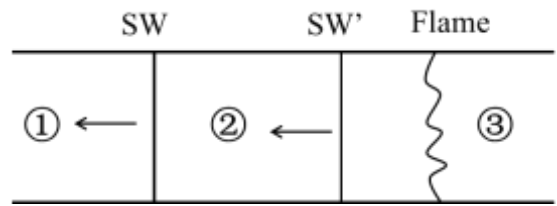

(a) superposition of shock waves propagating in the same direction

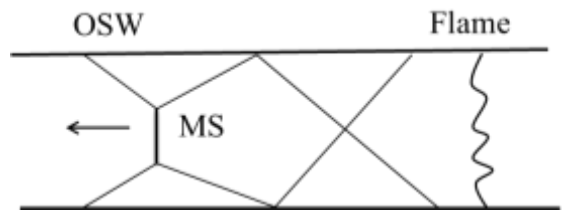

(b) Mach Stem formed by shock/shock interaction

Fig.1 Typical flow field structures of transient engine unstart process

This shock/shock interaction process is transient, nonlinear, discontinuous, and totally unpredictable. This transient and nonlinear shock/shock interaction process is very similar to the DDT transition process, which was discovered in detonation experiments more than one hundred years ago and is still a fundamental problem of combustion and detonation theory [27-32]. Most of the detonation initiation is initiated by the DDT process. In experiments, when a deflagration wave propagating with about 60% C-J detonation velocity ($D_{C-J}$) or close to the sound velocity of the combustion products, the DDT process occurs and a C-J detonation is triggered immediately and abruptly. This critical condition is called C-J deflagration, which is also the maximum propagating velocity of deflagration waves.

We conducted analysis about the mechanism of DDT process on the basis of shock/shock interaction and gave a theoretical equation to predict this critical condition [33]. The theoretical equation is given in Eq.(1).

$$\frac{6M_1^2}{M_1^2+5} = \frac{7M_{C-J}^2}{6}\frac{T_1}{T_0} \tag{1}$$

where, $M_1$ is the critical Mach number of the leading shock wave at DDT, $M_{C-J}$ the Mach number of C-J detonation in the initial detonable mixture, $T_0$ the total constant-volume combustion temperature and $T_1$ the initial static temperature, respectively. In this equation, $M_{C-J}$, $T_0$ and $T_1$ are constants when the gas properties and initial conditions are determined. Therefore, $M_1$ is also a constant for given detonable gas and initial condition. When the velocity of the supersonic combustion wave reaches the velocity of C-J deflagration wave, DDT process occur and a C-J detonation wave is formed immediately.

There are three important properties in Eq.(1). Firstly, the equality is also established when $M_1 = M_{C-J} = 0$. Secondly, there is a finite value for the C-J detonation Mach number $M_{C-J}$. The right hand should be smaller than 6, otherwise, the Mach number $M_1$ will become negative. Thirdly, the solution is $M_1 \approx M_{C-J} \approx 1$ when $T_1/T_0 = 1$, which means that the C-J detonation, the C-J deflagration and the sound wave become the same phenomena under this limit condition. These three important properties make sure that the Eq.(1) is physical.

Valiev et al. also put forth a semiempirical formula to predict the critical velocity of C-J deflagration at DDT [34,35]. This formula is given in Eq.(2). We can find from Eq.(2) that there is a constant factor between C-J deflagration velocity and C-J detonation velocity, which is a function of the specific heat ratio $\gamma$ of burnt gases.

$$(u_{C\text{-}J})_{\text{deflagration}} \approx \frac{\gamma(\gamma-1)+2(\gamma+1)}{2(\gamma+1)^2}(u_{C\text{-}J})_{\text{detonation}} \qquad (2)$$

The C-J detonation velocity, the sound velocity of detonation products and the velocity of C-J deflagration of H$_2$/air mixture at 0.1MPa, 300K and ER=1.0 calculated by Eq.(1) and Eq.(2) are plotted in Fig.2. The specific heat ratio $\gamma=1.36$ is assumed for Eq.(2). The agreement of these two equations is pretty good. The velocity of C-J deflagration is close to the sound velocity of the combustion products, which is about 40~60% D$_{C\text{-}J}$. This is also the maximum velocity of deflagration waves. When the combustion wave is accelerated to this critical value, DDT process will occur immediately. We think that the mechanism of engine unstart of scramjets is similar to the DDT process.

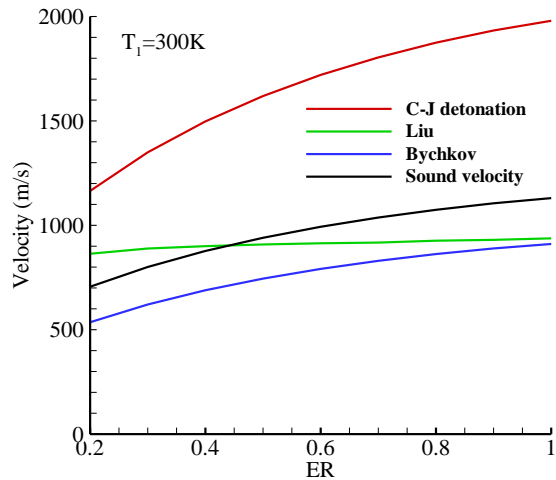

Fig.2 Comparison of C-J detonation velocity, sound velocity and C-J deflagration velocity of

H$_2$/Air mixture at 0.1MPa, 300K and ER=1.0

The velocity of C-J deflagration discussed in Fig.2 are the results at the initial static temperature of 300K. But the static temperature at the entrance of combustor is higher than 300K in scramjets, so the influence of initial static temperature on the DDT process should be considered. Eq.(1) contains the initial static temperature of $T_1$, so we use Eq.(1) to study the influence of initial temperature. The results of C-J detonation and C-J deflagration at initial temperature of 1000K and 1500K are plotted in Fig.3, respectively. The interesting thing is that the velocity of C-J deflagration is approaching the velocity of C-J detonation at higher initial temperature. This means that any value of deflagration velocity below $D_{C-J}$ is possible in scramjets. The reason is that the Mach number of shock wave becomes smaller at higher initial temperature because of the higher sound velocity although the velocity is still very fast. As a result, the nonlinear shock/shock interaction becomes weaker and produces many solutions.

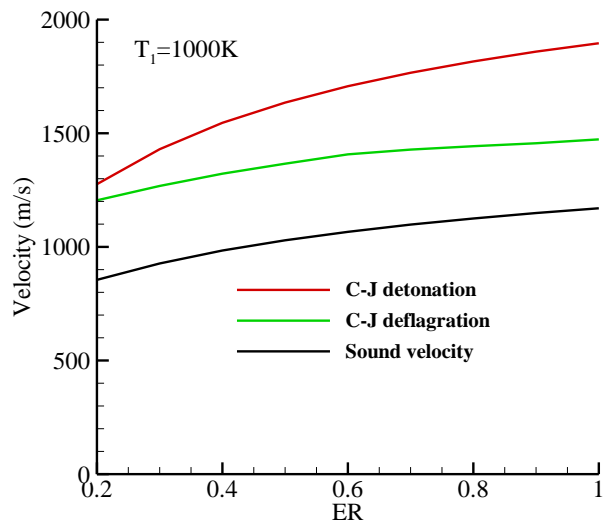

(a) $T_1$=1000K

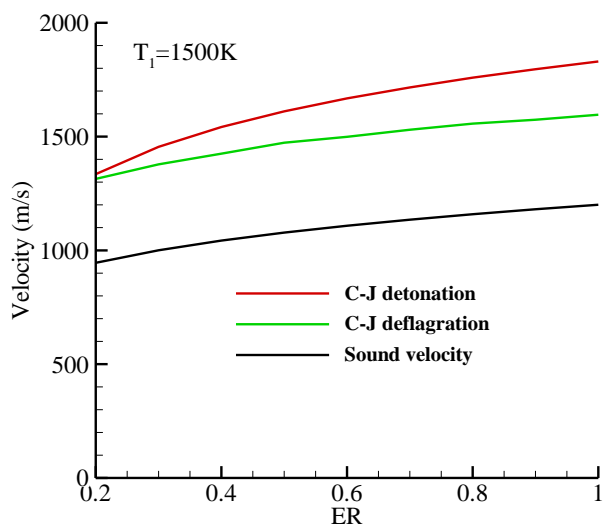

(b) $T_1$=1500K

Fig.3 C-J deflagration velocity of $H_2$/Air mixture at different initial temperature

From the above discussion, we find that the C-J deflagration is very unstable and the DDT process is always triggered when the propagation velocity of flame is close to the sound velocity of combustion products. This process is very difficult to control and should be avoided in the operation of scramjets. The engine unstart usually takes place in the scramjets operating in the flight Mach number region of Mach 5~8, where the velocity in the isolator is lower or close to the C-J detonation velocity. If the velocity in the isolator is faster than the C-J detonation velocity, the scramjets will operate stably. The experimental results of HIFiRE M12REST engine support this point of view. The C-J detonation is the strongest shock wave produced by combustion in theory, so we can define it as the stable operation boundary for scramjets.

This theory is demonstrated by many experimental results and the idea of this study is also inspired by these valuable experimental results. Table 1 gives the summary of some typical scramjets experimental results. We can find from Table 1 that the engine unstart or the pre-combustion shock wave always occur when the corresponding velocity of C-J detonation is close to the velocity in the isolator. Note that in the Case 4, the air-throttling method was used to promote the combustion, so the strength of the shock wave cannot be determined theoretically. In reality, the structures of the supersonic combustion flow field are very complicated including overdriven detonation, boundary layer separation, shock/shock interaction, Mach reflections, oblique shock waves, separation bubbles, and so on. This theory cannot cover all the details and it just gives a baseline principle for the stable operation of scramjets.

Table 1 Summary of some typical scramjets experimental results

| Cases | Fuel | Velocity in isolator (m/s) | Max ER without pre-combustion shock | $D_{C-J}$ under the same ER (m/s) | References |
|---|---|---|---|---|---|
| 1 | $H_2$ | 1720 | 0.50 | 1635 | [7][8][9][10] |
| 2 | $H_2$ | 1000 | 0.10 | 985 | [11] |
| 3 | $C_2H_4$ | 1000 | 0.32 | 1133 | [12][13] |
| 4 | $C_2H_4$ | 1060 | 0.39 | 1434 | [15][16] |
| 5 | $C_2H_4$ | 900 | 0.21 | 1139 | [19] |
| 6 | $H_2$ | 1750 | 0.48 | 1612 | [20][21][22] |
| 7 | $H_2$ | 2500 | 1.26 (stable) | 2039 | [23] |

**3 Propulsive Performance of Scramjets**

Once the C-J detonation is defined as the stable operation boundary of scramjets, we can use it to calculate the limiting propulsive performance of scramjets and discuss the influence of different key parameters. In the following part, the fuel of $H_2$ and hydrocarbon $CH_4$, $C_3H_8$, $C_8H_{18}$ are selected as representatives. The key parameters discussed are the static temperature at the entrance of combustor and the equivalate ratio. The C-J detonation velocity and pressure ratio (pressure-gain) are calculated by using C-J detonation theory, which are important parameters for determining the steadiness and thrust of scramjets. The Gaseq software are used to calculate the detonation parameters.

Figure 4 shows the C-J detonation velocity under different equivalence ratio at initial static temperature of 300K. From Fig.4 we can find that the equivalence ratio influences the C-J detonation velocity very much. The C-J detonation velocity is 1979m/s for $H_2$/air mixture at ER=1.0 and 1800 m/s for $C_3H_8$/air mixture at ER=1.0, respectively. The hydrocarbons have very close C-J detonation velocity. If the velocity in the isolator of scramjets is 1500m/s, the C-J detonation at ER=1.0 will propagate upstream and lead to the engine unstart. In order to decrease the C-J detonation velocity, we have to decrease the ER. According to Fig.4, if the velocity in the isolator is 1500m/s, the theoretical maximum equivalence ratio is ER=0.4 for $H_2$/air mixture and ER=0.5 for hydrocarbon/air mixture. For hydrocarbon/air mixture, the ignition delay time is longer and it is difficult to initiate a C-J detonation, so the ER can be a little larger than the theoretical value. But it is also very possible to form a stronger shock wave close to C-J detonation and the ER cannot be increased without limitation even without DDT.

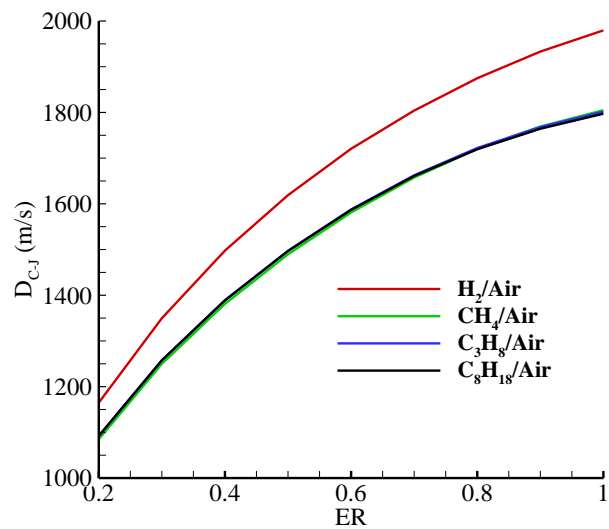

Fig.4 C-J detonation velocity under different ER at 300K

The influence of the static temperature on the C-J detonation velocity at ER=1.0 is shown in Fig.5. From Fig.5 we can see that the C-J detonation velocity is not very sensitive to the static temperature. For the $H_2$/air mixture at ER=1.0, the C-J detonation velocity is 1979m/s at 300K and 1830m/s at 1500K, respectively. For the hydrocarbon/air mixture at ER=1.0, the C-J detonation velocity is 1801m/s at 300K and 1736m/s at 1500K, respectively. The hydrocarbons also have very close C-J detonation velocity under different static temperature. The C-J detonation velocity of hydrocarbon/air mixture is about 100~200m/s slower than that of $H_2$/air mixture, which means that the combustion flow field of scramjets with hydrocarbon as a fuel is more stable than that of hydrogen.

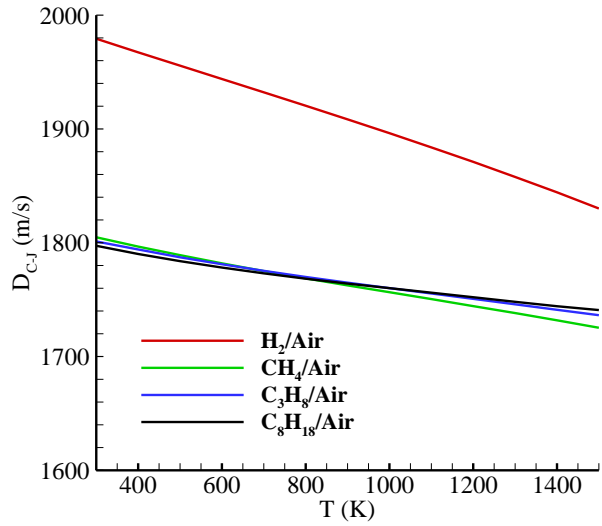

Fig.5 C-J detonation velocity under different static temperature at ER=1.0

The combustion pressure ratio or pressure-gain is a very important parameter to predict the thrust of scramjets. The pressure ratios of C-J detonation under different static temperature at ER=1.0 are plotted in Fig.6. The theoretical results show that the static temperature at the entrance of combustor influences the pressure ratio very much. For the $H_2$/air mixture, the pressure ratio is 15 at 300K and only 2.93 at 1500K, respectively. For the $C_3H_8$/air mixture, the pressure ratio is 17.48 at 300K and only 3.59 at 1500K, respectively. In order to increase the pressure ratio, the static temperature at the entrance of combustor should be decreased. These results also show that the pressure ratio of hydrocarbon/air is higher than that of hydrogen/air mixture, which means that hydrocarbon can produce larger thrust than hydrogen under the same condition.

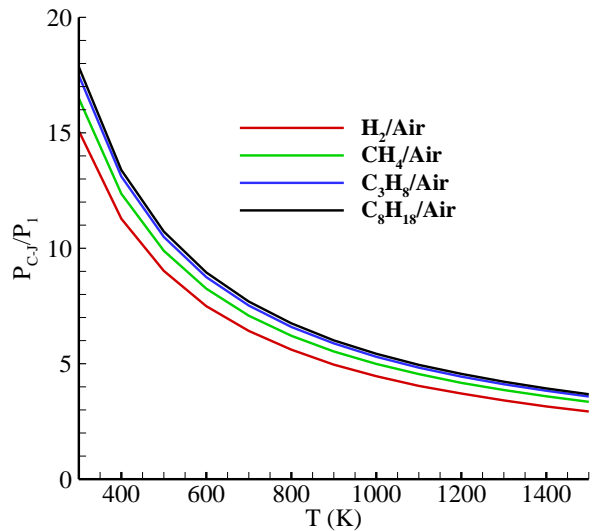

Fig.6 Pressure ratio under different static temperature at ER=1.0

The pressure ratio or pressure-gain is very sensitive to the static temperature. So, the influences of ER on pressure ratio are plotted in Fig.7 at initial temperature of 1000K and 1500K, respectively. These results have two important implications. Firstly, the difference between pressure ratio of $H_2$/air and hydrocarbons/air at lower ER is small, but it becomes very big at higher ER. The pressure ratio of $C_8H_{18}$/air at ER=1.0 is about 25% higher than that of $H_2$/air. This means that hydrocarbons can produce more thrust than hydrogen. Secondly, the slopes of these curves are very different. For the $H_2$/air mixture, the slope is higher at lower ER but smaller at higher ER. The curve of $H_2$/air becomes almost flat when ER>0.6. This means that we can obtain an overall optimized propulsive performance of hydrogen fueled scramjets at ER=0.5~0.6. But for the hydrocarbon/air mixtures, the pressure ratio increases with ER almost linearly and the maximum and optimized thrust is obtained at ER=1.0. Finally, we want to say that constant-volume combustion theory can also be used to predict the pressure ratio of scramjets qualitatively, but it cannot predict the velocity of the shock wave. This is the priority of C-J detonation theory.

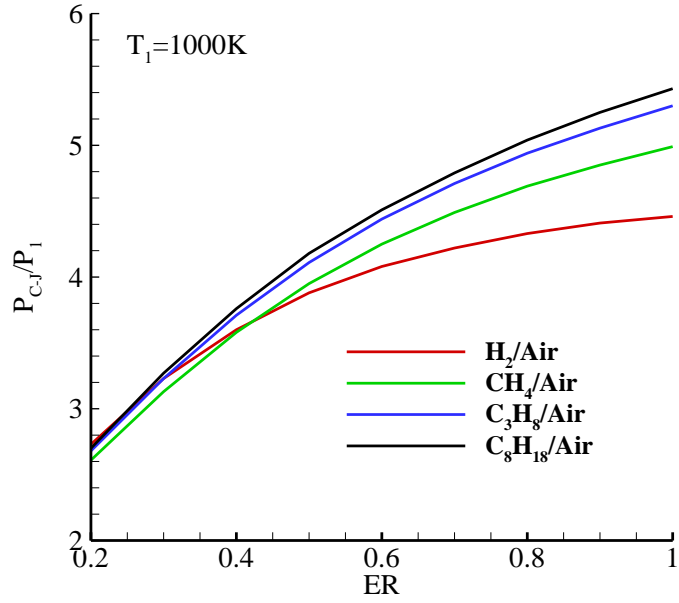

(a) pressure ratio at 1000K

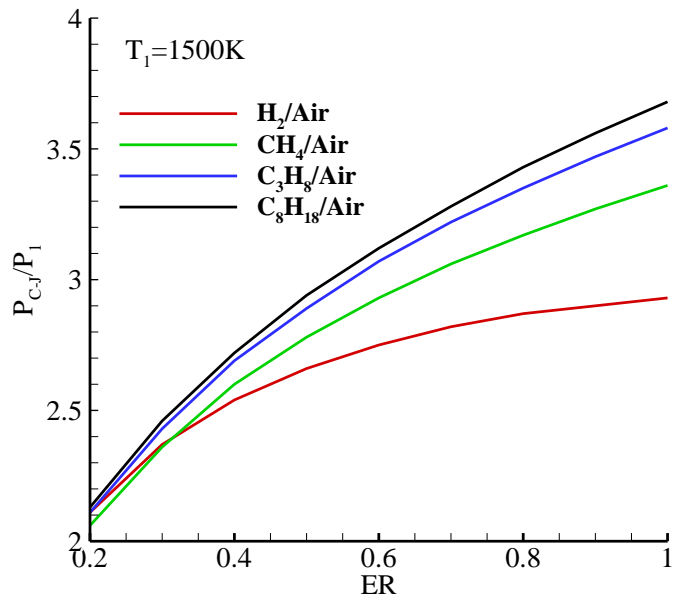

(b) pressure ratio at 1500K

Fig.7 Pressure ratio under different ER at 1000K and 1500K

**4 The method to increase the thrust of high Mach number scramjets**

According to the theoretical analysis and numerical simulations [36-38], the combustion flow field of scramjets is stable if the velocity of air in the isolator is greater than C-J detonation velocity at ER=1.0. Therefore, for the high flight Mach number scramjets with Ma≥10, additional fuel and oxygen can be added into the combustor to increase the thrust of scramjets as long as the velocity of the shock wave produced by the normal combustion, additional combustion and additional mass injection is slower than the velocity in the isolator.

The mechanism can be explained as follows. The scramjets can be considered as a shock tube. The high pressure and high temperature combustion products are the driver gas and the air in the isolator is the driven gas. The relationship between the driver pressure ratio and the shock wave Mach number is given in Eqs.(3)-(4).

$$P_{41} = \left[1 + \frac{2\gamma_1}{\gamma_1+1}(M_s^2 - 1)\right]\left[1 - \frac{\gamma_4-1}{\gamma_1+1}a_{14}\left(M_s - \frac{1}{M_s}\right)\right]^{-\frac{2\gamma_4}{\gamma_4-1}} \quad (3)$$

$$a_{14} = \sqrt{\frac{\gamma_1 M_4 T_1}{\gamma_4 M_1 T_4}} \quad (4)$$

where, $P_{41}$ is the pressure ratio of driver gas, $M_s$ the Mach number of the incident shock wave, $a_{14}$ the sound velocity ratio of the driven gas and driver gas, $\gamma_1$ and $\gamma_4$ the specific heat ratio of the driven gas and driver gas, $T_1$ and $T_4$ the static temperature of the driven and driver gas, $M_1$ and $M_4$ the molecular weights of driven and driver gas, respectively.

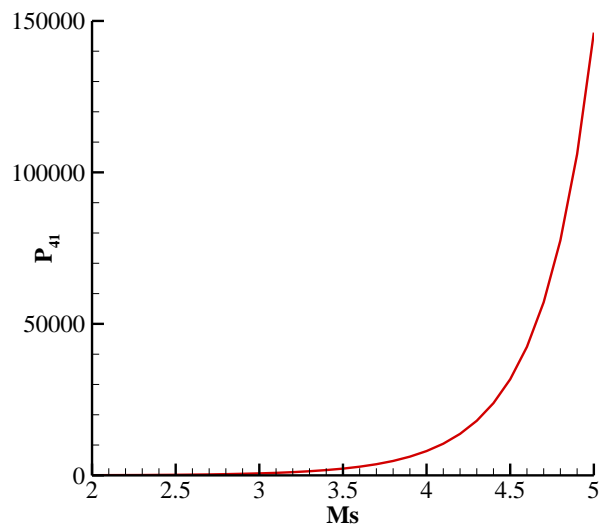

Fig.8 Relationship between driver pressure ratio and shock wave Mach number

Suppose the air static temperature in the isolator is 1000K and the driver gas is the high pressure and high temperature combustion products of hydrogen and oxygen at ER=1.0. The relationship between the Mach number and the pressure ratio calculated according to Eqs.(3)-(4) is plotted in Fig.8. We can see that the pressure ratio increases dramatically as the Mach number increases. The Mach number 4.0 in the isolator corresponds to the flight Mach number 10 and the Mach number 4.5 in the isolator corresponds to the flight Mach number 12. We can find that it is almost impossible for the combustion to produce enough back pressure to produce a shock wave in the isolator greater than Mach 4.5 without the consideration of boundary layer separation and reduction of engine flow passage aera. Therefore, extra fuel and oxygen can be injected into the scramjets combustor to increase its thrust for high Mach number scramjets. This looks like the combination of scramjets and rocket engines.

**5 Conclusions**

The mechanism of instability and engine unstart of scramjets are discussed theoretically in this paper. The point of view of shock/shock interaction and deflagration-to-detonation transition is introduced to explain the mechanism of this transient and nonlinear process. The theoretical analysis comes to the following conclusions.

(1) The intensive heat release of supersonic combustion produces shock waves and boundary separation. The shock/shock interaction or Mach reflection will form stronger oblique shock waves or normal shock waves. If the velocity of the pre-combustion shock wave is faster than the velocity in the isolator, it will propagate upstream and lead to the engine unstart.

(2) The C-J detonation is the strongest shock wave produced by combustion in theory and it is defined as the stable operation boundary of scramjets. The maximum equivalence ratio of scramjets is determined by the corresponding C-J detonation velocity, which should be slower than the velocity in the isolator. In this case, the combustion flow field of scramjets is stable.

(3) A theoretical equation is put forth to calculate the C-J deflagration velocity. The results show that the C-J deflagration velocity is very close to the C-J detonation velocity at higher initial static temperature. This means that any value of deflagration velocity below C-J detonation is possible even without DDT process in scramjets. Therefore, it is reasonable that the C-J detonation is defined as the stable operation boundary of scramjets

(4) The C-J detonation theory is used to study the influences of different key parameters on the limiting propulsive performance, including the static temperature, the equivalence ratio and the type of fuel. The detonation velocity is very sensitive to equivalence ratio, while the pressure ratio is very sensitive to the static temperature. The hydrocarbons are better than hydrogen because it can produce lower detonation velocity and higher combustion pressure ratio.

(5) For the high flight Mach number scramjets, the velocity in the isolator is faster than the C-J detonation velocity. Therefore, extra fuel and oxygen can be added into the combustor to increase the thrust further as long as the shock wave produced by combustion is slower than the air velocity in the isolator.

**Acknowledgments**